\documentclass[apj]{emulateapj}

\usepackage{natbib}
\usepackage{graphicx}

\shorttitle{The Origin of Multiple Nuclei in (U)LIRGs}
\shortauthors{Matsui et al.}

\begin{document}

\title{ORIGIN OF MULTIPLE NUCLEI IN ULTRALUMINOUS INFRARED GALAXIES}

\author{Hidenori Matsui\altaffilmark{1},
Takayuki R. Saitoh\altaffilmark{2,3},
Junichiro Makino\altaffilmark{1,2,3},
Keiichi Wada\altaffilmark{4},
Kohji Tomisaka\altaffilmark{1,2},
Eiichiro Kokubo\altaffilmark{1,2},
Hiroshi Daisaka\altaffilmark{5},
Takashi Okamoto\altaffilmark{6},
\&
Naoki Yoshida\altaffilmark{7}}

\altaffiltext{1}{Center for Computational Astrophysics, National Astronomical Observatory of Japan,
2-21-1 Osawa, Mitaka, Tokyo 181-8588}
\altaffiltext{2}{Division of Theoretical Astronomy, National Astronomical Observatory of Japan,
2-21-1 Osawa, Mitaka, Tokyo 181-8588}
\altaffiltext{3}{Interactive Research Center of Science, Tokyo Institute of Technology, 2-21-1, Ookayama, Meguro, Tokyo, 152-8551}
\altaffiltext{4}{Graduate School of Science and Engineering,  Kagoshima University, 
1-21-35, Korimoto, Kagoshima, 890-0065 
}
\altaffiltext{5}{Graduate School of Commerce and Management, Hitotsubashi University,
Naka 2-1 Kunitachi, Tokyo 186-8601}
\altaffiltext{6}{Center for Computational Sciences, University of Tsukuba,
1-1-1, Tennodai, Tsukuba, Ibaraki 305-8577, Japan}
\altaffiltext{7}{Institute for the Physics and Mathematics of the Universe, University of Tokyo,
5-1-5, Kashiwanoha, Kashiwa, Chiba, 277-8583, Japan}

\received{}
\accepted{}

\begin{abstract}
Ultraluminous infrared galaxies (ULIRGs) with multiple ($\ge 3$) nuclei are frequently observed.  It has been suggested that these nuclei are produced by multiple major mergers of galaxies. The expected rate of such mergers is, however, too low to reproduce the observed number of ULIRGs with multiple nuclei. We have performed high-resolution simulations of the merging of two gas-rich disk galaxies. We found that extremely massive and compact star clusters form from the strongly disturbed gas disks after the first or second encounter between the galaxies. The mass of such clusters reaches $\sim 10^8~M_{\odot}$, and their half-mass radii are $20-30~\rm{pc}$. Since these clusters consist of young stars, they appear to be several bright cores in the galactic central region ($\sim \rm{kpc}$). The peak luminosity of these clusters reaches $\sim 10~\%$ of the total luminosity of the merging galaxy. These massive and compact clusters are consistent with the characteristics of the observed multiple nuclei in ULIRGs. Multiple mergers are not necessary to explain multiple nuclei in ULIRGs.
\end{abstract}


\section{Introduction}

Ultraluminous and luminous infrared galaxies (ULIRGs/LIRGs) have 
strong infrared (IR) luminosities of $L_{\rm{IR}} > 10^{12}~L_{\odot}$
and $10^{11}~L_{\odot} \le L_{\rm{IR}} \le 10^{12}~L_{\odot}$,
respectively \citep{san96}. Their intense IR luminosities are mainly due to starburst activities, and the contribution of active galactic nuclei (AGN) to  IR luminosity seems to be rather limited, i.e., $\sim20~\%$ \citep{ris06,far07,nar08}. The starburst activities in (U)LIRGs are likely to be triggered by mergings of galaxies, in particular gas-rich galaxies, since observations suggest that (U)LIRGs have complex and disturbed morphologies \citep{san96}. In the $I$-band (F814W filter) images of 100 sampled (U)LIRGs at $0.05< z < 0.20$ taken with {\it Hubble Space Telescope} ({\it HST}), about $20~\%$ of  ULIRGs have multiple ($\ge 3$) nuclei \citep{bor00,cui01}. The fraction  increases to more than $80~\%$, if all probable galaxies are included  \citep{bor00}.

The origin of the multiple nuclei has been thought to be multiple major merger events. In order to explain multiple mergers, \citet{bor00} argued that the progenitors of (U)LIRGs, were compact groups of galaxies. However, the evolution timescale of compact groups is very long ($\sim H_0^{-1}$) \citep{ath97}. Therefore, the probability that two galaxies merge in a compact group and the merging galaxy has double cores is $p_{\rm pair} \sim t_{\rm{merg}}/H_0^{-1}$ $\sim 100~\rm{Myr}/10~{Gyr}$ $\sim 0.01$, where $t_{\rm{merg}}$ is the merging timescale of the galactic cores. The probability that another galaxy merges with such merging galaxy in the compact group and the merging galaxy has triple cores is $p_{\rm multi} = p_{\rm pair} \times \frac{1}{N_{\rm g}-2} \times p_{\rm pair} \sim 3 \times 10^{-5}$, where $N_{\rm{g}}$ ($\sim 5$) is the typical number of galaxies in a compact group \citep{deo91}. Then, the number density of ULIRGs with multiple cores is $\phi _{\rm{g}} \times p_{\rm multi}$ $\sim 3 \times 10^{-9}~\rm{Mpc}^{-3}$, where $\phi _{\rm{g}}$ ($\sim 10^{-4}~\rm{Mpc}^{-3}$) is the number density of galaixes in a compact group \citep{deo91,rib94}. Since the number density of ULIRGs is $10^{-7}~\rm{Mpc}^{-3}$ \citep{san03}, the fraction of multiple merger to ULIRGs is $(3 \times 10^{-9})/10^{-7} \sim 0.03$. Thus, it is unliky that multiple merger is a major cause of the observed multiple nuclei in (U)LIRGS. An alternative explanation for the origin of multiple nuclei in (U)LIRGs seems necessary.

Previous numerical simulations of merging galaxies \citep{mih92,mih96,bar96,kaz05,cox06,dim07,nar09,cov10} have shown that starbursts occurred during the merging process. However, these simulations failed to reproduce formation of star clusters as observed in many interacting and merging galaxies \citep{whi95,men08}, since their spatial and mass resolutions ({\it e.g.}, $100~\rm{pc}$ and $\sim 10^6~M_{\odot}$) were not sufficiently high to distinguish star forming region and formation of star clusters whose sizes are less than $100~\rm{pc}$. In addition, previous simulations did not allow the interstellar medium (ISM) to cool below $\sim 10^4~\rm{K}$. This is another reason why formation of star clusters was not reproduced in the previous simulations.

In order to reproduce formation of star clusters, high resolution simulations have been attempted. \citet{li04} have performed the simulation of a merging galaxy using Tree+SPH code GADGET. In this simulation, mass and spatial resolutions were $10~\rm{pc}$ and $6.6\times 10^3~M_{\odot}$, respectively. They assumed an isothermal ISM and used sink particles, that absorb their surrounding gas, to represent clusters. They have shown that a number of massive clusters form in the merging process. In this simulation, the most massive cluster has mass of $7.8 \times 10^7~M_{\odot}$. \citet{bou08} have performed the simulations of merging galaxies using sticky particles, that collide inelastically, instead of solving hydrodynamics. In these simulations, mass and spatial resolutions were $32~\rm{pc}$ and $7\times 10^3~M_{\odot}$, respectively. They have also shown formations of massive star clusters with masses of $10^{5-7}~M_{\odot}$. These simulations are, however, unrealistic in a sense that dynamical evolution of star clusters themselves cannot be properly followed due to the limitation of the sticky and sink particle methods. \citet{sai09a} improved the spatial and mass resolutions ($5$-$20~\rm{pc}$ and $10^{3-4}~M_{\odot}$) and the ISM model that is allowed to cool to $10~\rm{K}$. These simulations showed that the behavior of the multiphase ISM in the merging galaxies is considerably altered and the formation of shock-induced star clusters is naturally reproduced \citep{sai09a,sai11}. There are several high resolution simulations of merging galaxies, which resolve the low temperature gas ($300$-$500~\rm{K}$) using adaptive mesh refinement (AMR) methods \citep{kim09,tey10}. These simulations also showed the difference in the behavior of the multiphase ISM from that of ISM used in the previous simulations. \citet{kim09} considered the merger of low-mass galaxies ($\sim 1.8\times 10^10~M_{\odot}$). Several spiky peaks of star formation rate were seen in thier simulation although the most prominet starburst was found at the beginning the simualtions and the significant fraction of the ISM was blown away by energetic wind before the merging event. Thus, it would be hard to investigate detailed evolution of the ISM in merging galaxies. In \citet{tey10}, the polytropic equation of state was used instead of solving the energy equation, which is essentially different from \citet{sai09a} and present paper. The validity of this approximation is unclear for understanding the evolution of the ISM in the merging galaxies.

High-resolution images of the local (U)LIRGs obtained by the integral field spectroscopy using {\it William Herschel Telescope} \citep{gar09b,gar09a} and {\it VLT-VIMOS} \citep{alo09,alo10,mon10,Rod10} showed that (U)LIRGs generally have very complex structures, such as H$\alpha$ bright knots, rings, and tidal tails. As in the case of formation of star clusters, the reason why these structures have not been reproduced in numerical simulations might be just the inadequate treatment of ISM and limited resolution. Thus, high resolution simulations resolving multiphase ISM are essential to comprehend the complex structures in (U)LIRGs.

We have performed high-resolution simulations of merging galaxies with sufficiently high spatial resolution and cooling function of ISM that covers a wide range of temperature ($10~\rm{K}<T<10^{8}~\rm{K}$). In this paper, we focus on the origin of the observed ULIRGs with multiple nuclei. The detailed evolution of merging galaxies will be described in forthcoming papers.

The structure of this paper is as follows. We first describe the numerical methods and models in \S \ref{sec:Methods}. Numerical results are shown in \S \ref{sec:Results}. In \S \ref{sec:Discussion}, we compare our results with observations.  Conclusions and discussion are presented in \S \ref{sec:Conclusions}.

\section{Methods} \label{sec:Methods}

\subsection{Simulation Setup}

The model parameters of the initial disk galaxy are the same as those used in \citet{sai09a}.  The masses of the dark matter halo and old stellar and gas disks are $1.1\times 10^{11}~M_{\odot}$, $5.1\times 10^{9}~M_{\odot}$, and $1.2\times 10^9~M_{\odot}$, respectively. The gas fraction in the disk is $\sim 20$\% of the total disk mass. The gas fraction is set to be slightly higher than that of local spiral galaxies since the gas fraction decreases due to star formations during the isolated phase for $1~\rm{Gyr}$ before interactions  (see below). We also assume the halo gas component of which total mass is $1.1\times 10^{9}~M_{\odot}$. It has the same distribution as the dark matter halo. The initial gas temperatures are $10^{4}~\rm{K}$ in the disk and $10^{5}~\rm{K}$ in the halo. The scale radii of the stellar and gas disks are $4~\rm{kpc}$ and $8~\rm{kpc}$, respectively. The dark-matter halo and the stellar disk are expressed by $N$-body particles, whereas the gas disk is expressed by smoothed particle hydrodynamics (SPH) particles.

Unlike \citet{sai09a}, before starting the simulations, we let an isolated disk galaxy evolve for $1~\rm{Gyr}$ in order to stabilize the gas component. After this evolution, the amount of disk and halo gas in each galaxy is $1.8\times 10^9~M_{\odot}$. The SFR during this period is $\sim 0.5~M_{\odot}~\rm{yr}^{-1}$. During the first $1~\rm{Gyr}$, the disk galaxy holds a quasi-steady state and is free from a global instability due to gravity. The evolution of the isolated disk is different from that of a gas rich disk at high-$z$. Simulations of the gas-rich disk, in which gas mass fraction of a disk is $\ge 50$ \%, have shown that massive clumps form from gravitational instabillity of the gas rich disk \citep{nog99,imm04,bou07}. In contrast to the simulations of merging of such clumpy disks at high-z by \citet{bou11}, we focus on the merging process of two local disk galaxies.

We let the pre-evolved two galaxies collide in a parabolic orbit with the pericentric distance of $7.5~{\rm kpc}$. The simulations start with the initial separation  of $75~{\rm kpc}$, where the separation distance is measured between the mass centers of the two galaxies.

We have performed nine runs. Three of them are high-resolution runs ($H_{\rm{XX}}$ in Tab.~\ref{param_tab}) and the rests are low-resolultion runs ($L_{\rm{XX}}$). The initial masses of SPH, star, and dark matter particles are the same. They are $7.5 \times 10^3~M_{\odot}$ and $3\times 10^4~M_{\odot}$, for $H_{\rm{XX}}$ and $L_{\rm{XX}}$ runs, respectively. Subscripts P, T, and R denote prograde, tilted, and retrograde spin axes of two galaxies. We performed six possible combinations of the spin orientations in low resolution and only PP and TT cases in high resolution. The gravitational softening length, $\epsilon$, is set to be $20~\rm{pc}$ for eight runs except for $H_{\rm{TT},5\rm{pc}}$ run in which the softening length is $5~\rm{pc}$.

The orientation of the disk axis is specified by two angles, $i$ and $\omega$. Here, $i$ is the inclination and $\omega$ is the argument of pericenter. They are defined as follows. We use the coordinate system in which the orbital plane of two galaxies is in the $x$-$y$ plane. The Laplace-Runge-Lenz vector of the orbit is $\mbox{\boldmath $e$}_{\rm{p}}=(1,0,0)$. The spin axis of a galaxy is then given by $\mbox{\boldmath $e$}_{\rm{g}}=(\sin \omega \sin i, \cos \omega \sin i,\cos i)$. These angles are defined relative to the orbit of each galaxy. Thus, in the case of ``TT'' runs, spin axes of two galaxies are parallel to each other.

\begin{table*}
\begin{center}
\caption{The model parameters.\label{param_tab}}
\begin{tabular}{lccccccccc}
\tableline\tableline
Run & $i_{1}^{\rm a}~\rm{[deg]}$ & $\omega _{1}^{\rm b}~\rm{[deg]}$ & $i_{2}^{\rm c}~\rm{[deg]}$ & $\omega _{2}^{\rm d}~\rm{[deg]}$ & $m^{\rm e}~[M_{\odot}]$
& $N^{\rm f}_{\rm{SPH}}$ & $N^{\rm g}_{\rm{star}}$ & $N^{\rm h}_{\rm{DM}}$ & $\epsilon^{\rm i}~[\rm{pc}]$ \\
\hline
$H_{\rm{PP}}$	  & $0$     & - & $0$ & - & $7.5 \times 10^3$ & $514476$ & $1841042$ & $27720000$ & 20 \\
$H_{\rm{TT}}$	 & $-109$ & $-30$ & $71$ & $-30$ & $7.5 \times 10^3$ & $514476$ & $1841042$ & $27720000$ & 20 \\
$L_{\rm{PP}}$   & $0$ & - & $0$ & - & $3.0 \times 10^4$ & $133696$ & $447994$ & $6930000$ & 20 \\
$L_{\rm{PR}}$   & $0$ & - & $180$ & - & $3.0 \times 10^4$ & $133696$ & $447994$ & $6930000$ & 20 \\
$L_{\rm{RR}}$   & $180$ & - & $180$ & - & $3.0 \times 10^4$ & $133696$ & $447994$ & $6930000$ & 20 \\
$L_{\rm{PT}}$  & $0$ & - & $71$ & $30$ & $3.0 \times 10^4$ & $133696$ & $447994$ & $6930000$ & 20 \\
$L_{\rm{TR}}$   & $-109$ & $30$ & $180$ & - & $3.0 \times 10^4$ & $133696$ & $447994$ & $6930000$ & 20 \\
$L_{\rm{TT}}$  & $-109$ & $-30$ & $71$ & $-30$ & $3.0 \times 10^4$ & $133696$ & $447994$ & $6930000$ & 20 \\
$H_{\rm{TT},5\rm{pc}}$	 & $-109$ & $-30$ & $71$ & $-30$ & $7.5 \times 10^3$ & $514476$ & $1841042$ & $27720000$ & 5\\
\tableline
\tableline
\end{tabular}
\end{center}
$^{\rm a}$ The disk inclination for the galaxy $1$. 
$^{\rm b}$ The argument of the pericenter for the galaxy $1$. 
$^{\rm c}$ The disk inclination for the galaxy $2$. 
$^{\rm d}$ The argument of the pericenter for the galaxy $2$. 
$^{\rm e}$ The mass of SPH, star, and dark matter particles.
$^{\rm f}$  The number of SPH particles.
$^{\rm g}$ The number of star particles.
$^{\rm h}$ The number of dark matter particles.
$^{\rm i}$ The gravitational softening length.
\end{table*}

\subsection{Numerical Method} \label{sec:NumericalMethod}

Numerical simulations were performed by an $N$-body/SPH code {\tt ASURA} \citep{sai09a} that can optionally utilize GRAPE \citep{sug90} in order to accelerate calculation of gravitational force.  We adopted a time-step limiter \citep{sai09b} for the time-integration of SPH particles with individual time-steps, which keeps the differences in time-steps of neighboring particles to be small enough to handle the strong shocks correctly.  We also used the FAST scheme \citep{sai10} that allows self-gravitating SPH particles to use  different time-steps for integrations of hydrodynamics and gravity. This method accelerates simulations without losing the accuracy of the time-integration.  The leapfrog method is used for the time integrations of both gravity and hydrodynamics. These improvements allow us to follow the formation of cold and dense gas clumps and their expansion by supernovae (SNe) feedback without numerical problems.

For the radiative cooling and heating, we follow the treatment of \citet{sai08}, in which the gas is allowed to cool down to $10~\rm{K}$. The treatment of star formation and SN feedback are also the same as those in \citet{sai09a}. When an SPH particle satisfied (1) high number density ($n_{\rm{H}}>100~\rm{cm}^{-3}$), (2) a low temperature ($T<100~\rm{K}$), and (3) a collapsing region ($\nabla \cdot \mbox{\boldmath$v$}<0$), the SPH particle spawns a star particle of which mass is one-third of the original SPH particle mass. We assume simple stellar polulation approximation for newly formed star particles. Salpeter's initial mass function with mass range of $0.1-100~M_{\odot}$ is adopted \citep{sal55}. Stars with mass $\ge 8~M_{\odot}$ in each star particle explode as Type II SNe and release $10^{51}~\rm{erg}$ of thermal energy per one SN into the surrounding ISM.

\section{Multiple Nuclei in ULIRGs} \label{sec:Results}

Initially, the two galaxies move on the parabolic orbits and approach to each other. At around $t=450~\rm{Myr}$, they reach the pericenter (first encounter). Since orbital angular momenta of their main bodies are converted into the disk internal spin, they lose their orbital angular momenta \citep{bar92b,her92,her93,mih96} and their trajectories deviate from the parabolic orbits. The main bodies approach to each other again (second encounter). Finally, after several encounters, their cores completely merge.

\begin{figure*}
\epsscale{0.8}
\plotone{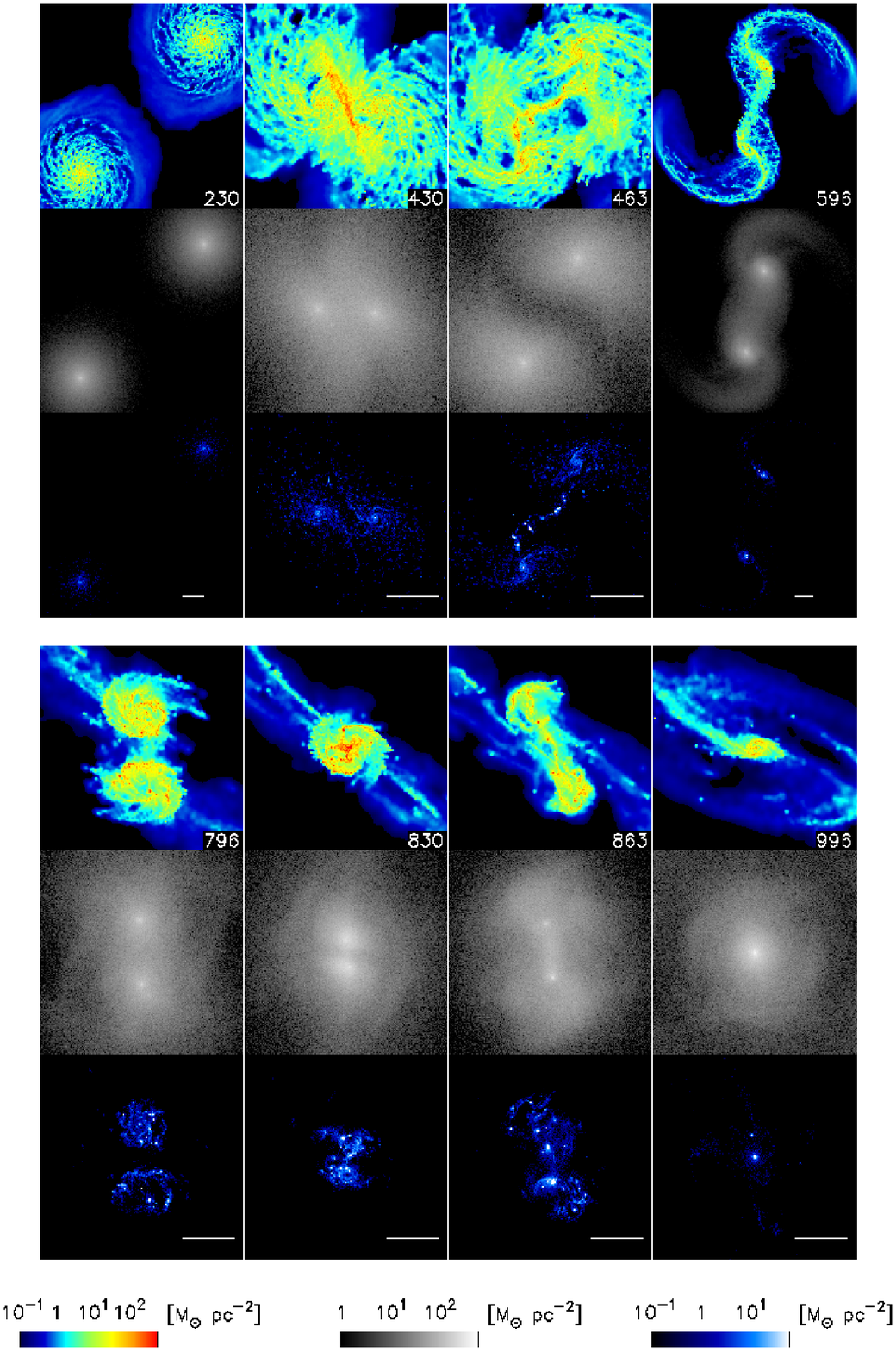}
\caption{
Snapshots of run $H_{\rm{PP}}$. The upper color, middle gray-scale, and bottom color panels show surface density of gas, old stars, and newly formed stars that are born after $t=0$, respectively. In each upper panel, the number in the right-bottom corner displays simulation time (the unit is Myr). The white line in the right-bottom corner of each panel shows length of $5~\rm{kpc}$.
\label{snapshot_Hpp}}
\end{figure*}

\begin{figure*}
\epsscale{0.8}
\plotone{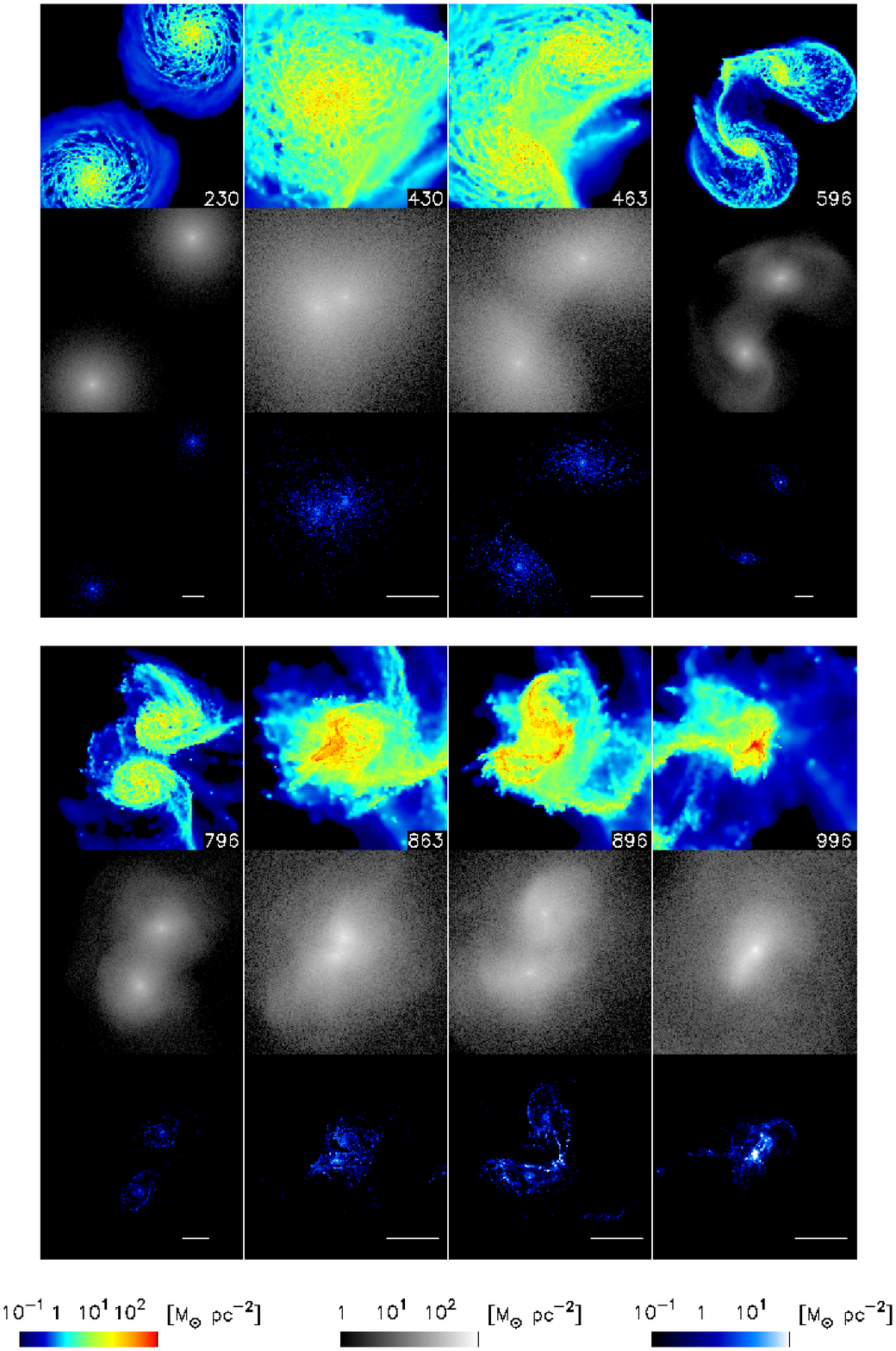}
\caption{
Same as Fig.~\ref{snapshot_Hpp} but run $H_{\rm{TT}}$.
The snapshots are projected to the disk plane.
\label{snapshot_Htt}}
\end{figure*}

Figures~\ref{snapshot_Hpp} and \ref{snapshot_Htt} show the evolution of distribution of gas and stellar particles for runs $H_{\rm{PP}}$ and $H_{\rm{TT}}$. Gas disks are strongly disturbed at the first encounter (the panels at $t=430$, $463$, and $596~\rm{Myr}$) and at the second encounter (the panels at $t=830$ and $863~\rm{Myr}$) in run $H_{\rm{PP}}$ and at the second encounter (the panels at $t=863$ and $896~\rm{Myr}$) in run $H_{\rm{TT}}$. Because of the strong perturbation, gas is compressed by large-scale shocks. The dense gas radiatively cools and becomes gravitationally unstable. As a result, massive star clusters form \citep[see also][]{sai10b,sai11}.

\begin{figure*}
\epsscale{1.0}
\plotone{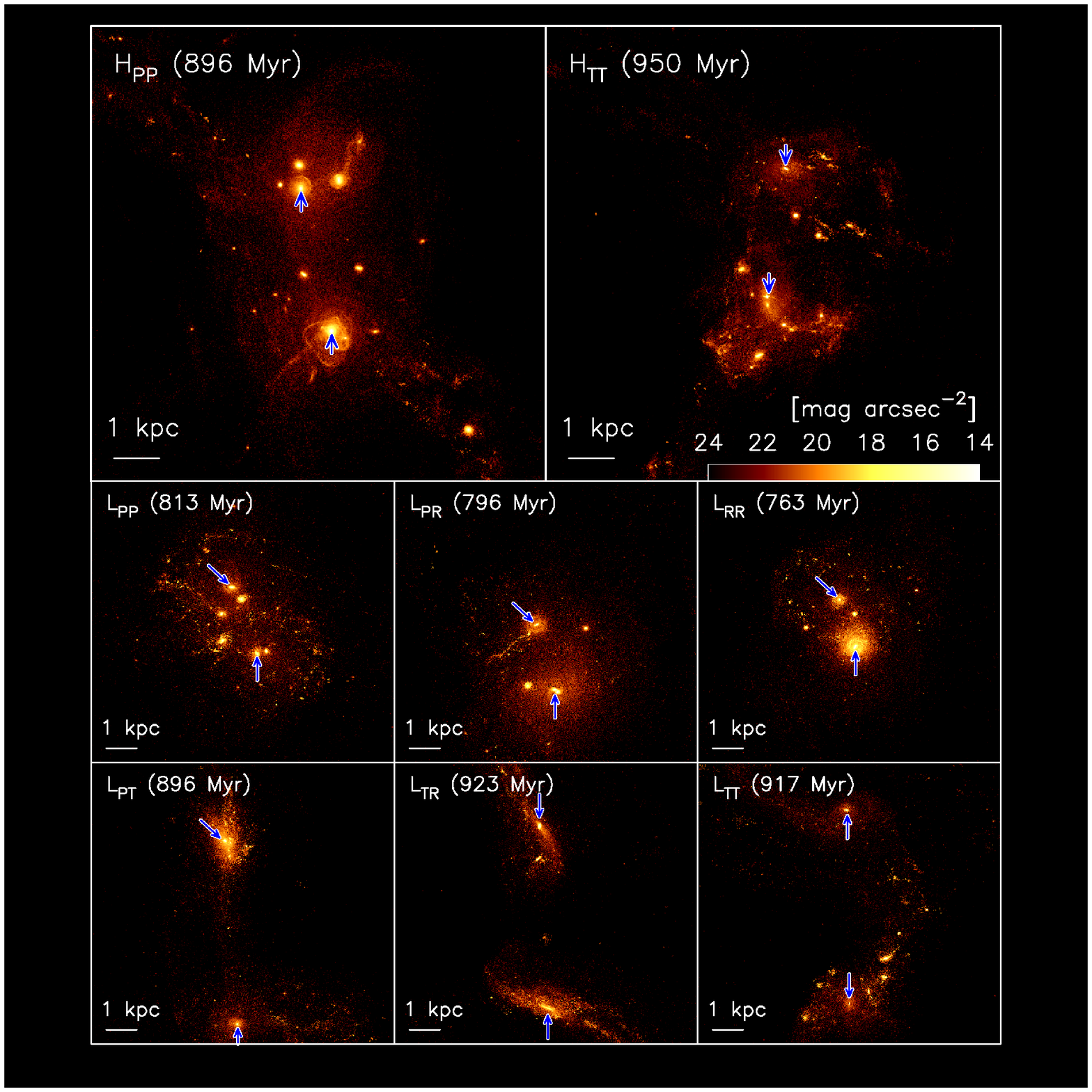}
\caption{
Simulated {\it HST} WFPC2 $I$-band (F814W filter) images from our simulations.  The color represents magnitude per square arcsecs.  The left and right panels in the top row show the results of runs $H_{\rm{PP}}$ and $H_{\rm{TT}}$, respectively.  These two runs are simulated with high resolution.  Six panels in the bottom two rows show the results of low resolution runs. We ignore the effect of dust extinction (see \S 4.2.).
\label{image}}
\end{figure*}

Figure~\ref{image} shows the synthesized $I$-band images from our merger
simulations after the massive star clusters formed. Here, the $I$-band
luminosity emitted from newly formed stars was calculated using population
synthesis code {\tt P\'EGASE} \citep{fio99}, and effects of dust absorption and re-emission were neglected. The two top panels show the high resolution runs, and the six lower panels show the low resolution runs. In all images, multiple compact sources appear in the central regions of galaxies. Two of them with arrows are the progenitor galactic cores, and the others are newly-formed massive star clusters. The absolute magnitudes of these clusters in $I$-band are $M_{I}\lesssim -17.0$, which are comparable to those of the galactic central cores. The effect of dust extinction is discussed in \S 4.2.

Since the cores form through the gravitational instability of strongly perturbed gas disks, they do not contain dark matter and consist only of young stars. On the other hand, the galactic luminous nuclei contain dark matter and a large amount of old stars that have formed before the merger event.  The properties of newly formed cores are, therefore, different from those of the galactic nuclei.  In addition, such cores are more massive than usual star clusters with $10^{6-7}~M_{\odot}$, by an order of magnitude. We, therefore, call the newly formed cores ``hypermassive star clusters''. Since the hypermassive star clusters form in the galactic central region ($\sim \rm{kpc}$), they are different from tidal dwarves that are formed in tidal arms \citep[e.g.,][]{bar92,bou08} .

The reason why they are extremely massive can be explained as follows. The total mass of the gas which becomes gravitationally unstable is very large ($\sim 1.0\times 10^9~M_{\odot}$) because of the strong large-scale gravitational and hydrodynamical disturbances from another galaxy at the encounter phase. The strong disturbances deform the gravitational potential into strongly non-axisymmentric shape.  Therefore, the rotation of the disk becomes ineffective in stabilizing long-wavelength perturbations.  Even though the gas disk fragments to a number of small clumps with $\sim 10^{6-7} M_{\odot}$, these small clumps are still gravitationally bound to each other and eventually merge to form larger clusters.

\begin{figure*}
\epsscale{1.0}
\plotone{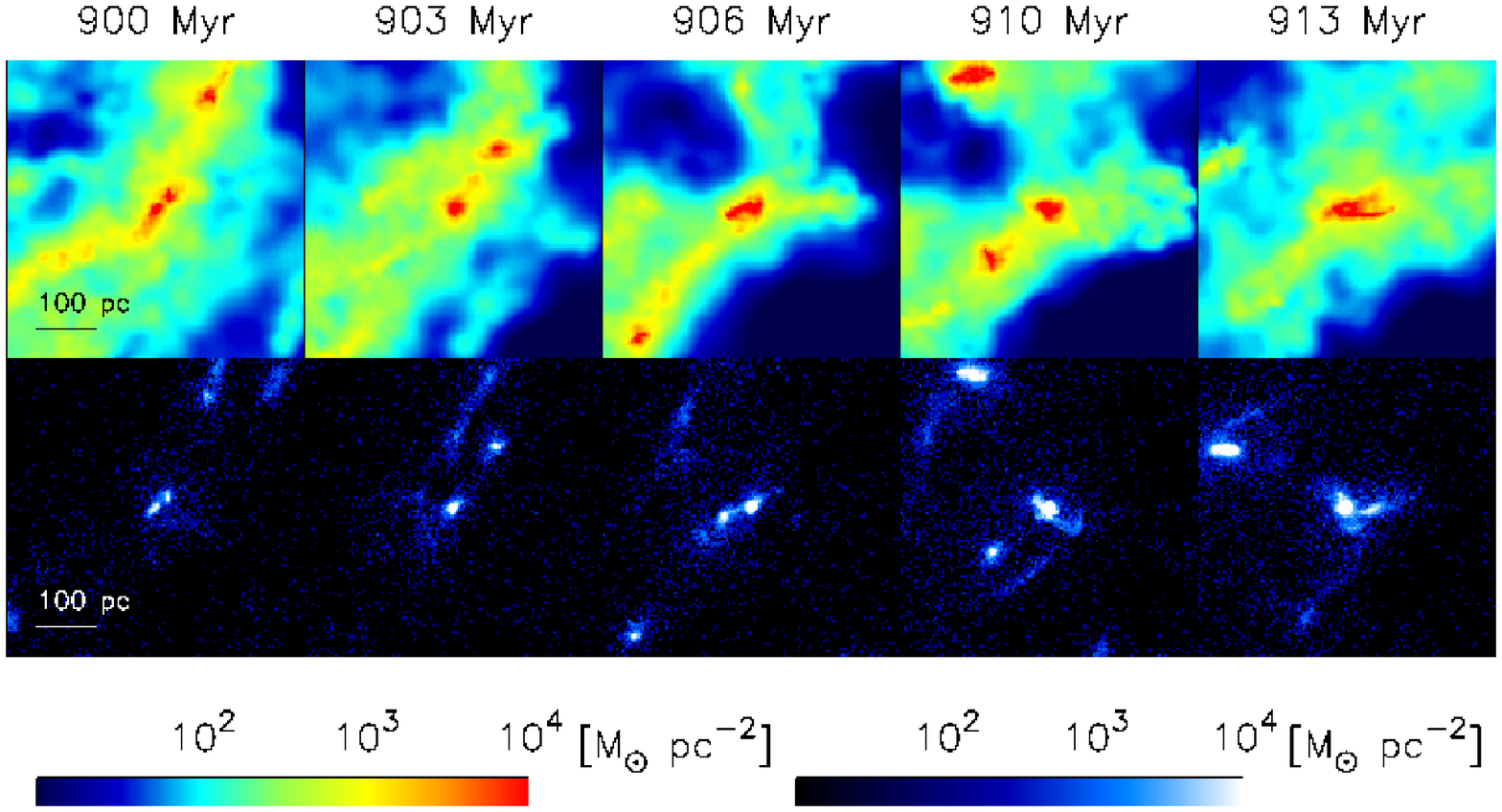}
\caption{
The formation of the most massive star cluster in run $H_{\rm{TT}}$ (the black curve in the right panels of Fig. \ref{mass_evolution}).  The upper panels show the evolution of the gas surface density while bottom ones display that of the stellar surface density.  The size of each panel is $500~\rm{pc}$ by $ 500~\rm{pc}$, and its center is taken as the center of mass of the main progenitor of the final cluster. Two clusters in the center of the panel at $900~\rm{Myr}$ has merged at $903~\rm{Myr}$. Two clusters at $903~\rm{Myr}$ approach and merge at $906~\rm{Myr}$ each other. The cluster in the left-bottom corner of the panel at $906~\rm{Myr}$ merges with the cluster at the center by $913~\rm{Myr}$.
\label{snapshot}}
\end{figure*}

Figure~\ref{snapshot} shows the snapshots during the formation of the most massive cluster from $900~\rm{Myr}$ to $913~\rm{Myr}$ in run $H_{\rm{TT}}$. Multiple gas clumps formed from the disturbed gas disk merge with each other  and become one massive cluster. In other words, the merging of smaller clumps within the short timescale causes quick growth of the mass of star clusters. As a result, hypermassive star clusters form.

\begin{figure*}
\epsscale{1.0}
\plotone{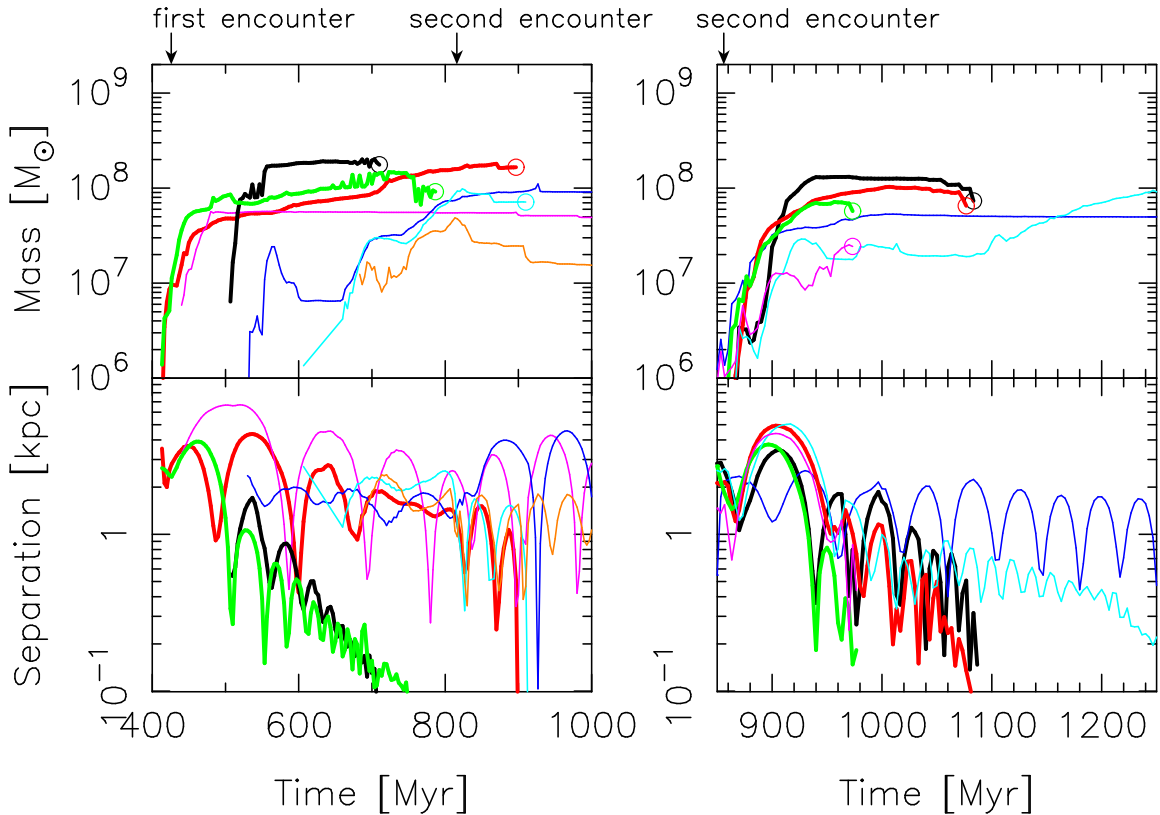}
\caption{
Time evolution of masses of hypermassive star clusters (upper panels) and
their distances from the galactic center (lower panels).  The galactic
center is given by the nearest one to the cluster before the coalescence of
galactic cores and by the galactic center of a merger remnant after that.
The left and right panels show the result of runs $H_{\rm{PP}}$ and $H_{\rm{TT}}$, respectively.  The three most massive star clusters are shown by the thick curves, and the others are shown by the thin curves.  We show the evolution of each clusters until it sinks to the galactic center (the distance becomes less than $100~\rm{pc}$) or it is disrupted by tidal forces \citep{sai06}. 
Once they sink to the galactic center less than $100~{\rm pc}$ or fully
disrupted by the tidal force, we put circles at these points and stop 
tracing their evolutions.
\label{mass_evolution}}
\end{figure*}

In Fig.~\ref{mass_evolution}, we show the evolution of the masses of these hypermassive star clusters (upper panels), and their distances from the center of the galaxy (lower panels). Here, the definition of the distance is the smaller of the two distances from the two galactic centers. The clusters first grow quickly, reaching about a half of the final mass in around $20~\rm{Myr}$. After that, the growth of the mass slows down, and the clusters fall to the center of the galaxy due to the dynamical friction.  As a result, the merger remnant has a very compact and luminous core similar to those observed in a large number of elliptical galaxies \citep[e.g.,][]{kor09}.

\begin{figure}
\epsscale{1.0}
\plotone{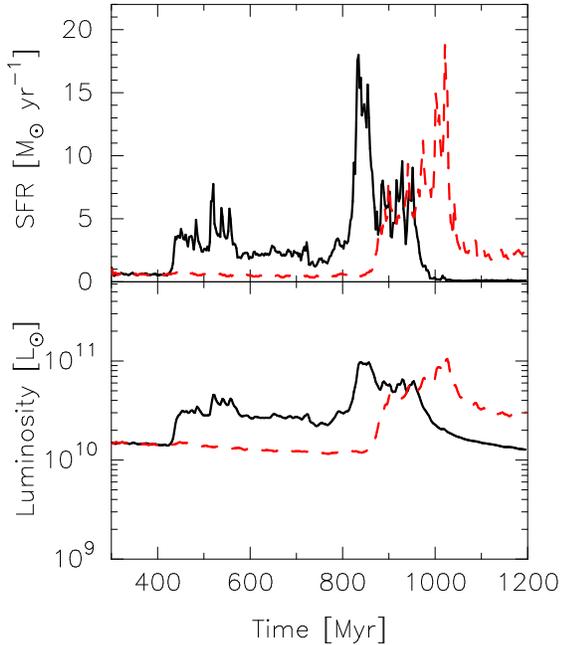}
\caption{
Time evolution of the SFR (upper panel) and the bolometric luminosity (lower panel) of merging galaxies. The black solid and red dashed curves show results of runs $H_{\rm{PP}}$ and $H_{\rm{TT}}$, respectively.  We show the mean quantities of every $1~\rm{Myr}$.
\label{sfr}}
\end{figure}

Figure~\ref{sfr} shows the evolution of the SFR and bolometric luminosity coming from the stars in the merging event. The effects of dust absorption and re-emission are not taken into account. If we take them into account, the IR luminosity would become higher since ultraviolet and optical photons are absorbed by dust and re-emitted in the IR regime. Therefore, we expect that the bolometric luminosity of our runs is a good indicator of the IR luminosity. In all runs, the maximum SFR reaches $20~M_{\odot}\rm{yr}^{-1}$, and the luminosity reaches $10^{11}~L_{\odot}$, which is about an order of magnitude higher than that before the encounter. The figure shows that the evolution of bolometric luminosity reflects the SFR. The peak luminosity is comparable to those of LIRGs \citep{san96}. If the progenitor galaxies are more massive and/or the initial gas fraction is much higher than those used in this simulation, the peak of SFR would become higher and the merger galaxy would become ULIRGs with infrared emission $\ge 10^{12}L_{\odot}$. AGNs would also provide an additional IR luminosity, although our simulations do not take into account the effect of AGNs. The influence of AGNs to the total IR luminosity is typically $\sim 20$ \% of the total one \citep{ris06,far07,nar08}. The peak mass-to-light ratio of our run, $M/L$, is less than $0.1$ and is comparable to that of ULIRGs \citep{col05,hin06}. The period of the high luminosity phase is $\sim 200~\rm{Myr}$. The hypermassive star clusters after their formation contain a large amount of dusty gas and young stars as shown in Fig.~\ref{snapshot}. Therefore, a strong infrared emission from the hypermassive star clusters is likely to be observable.

In order to study the dependence on the numerical resolution of the formation of hypermassive star clusters, we have performed simulations with different mass resolutions. In both high and low resolution runs, hypermassive star clusters with $\sim 10^{8}~M_{\odot}$ are formed as shown in Fig.~\ref{image}. Their formation process is the same as that of the high resolution runs. In addition, the time evolution of SFR is also similar to that of high resolution runs. In both runs, the peak of SFR is $20-30~M_{\odot}~\rm{yr}^{-1}$, and the duration time of the active star formation is about $200~\rm{Myr}$.

\section{Comparison with Observations} \label{sec:Discussion}

\subsection{Surface Brightness and Spatial Distribution of Multiple Nuclei}

Here, we compare our numerical results with observations of ULIRGs with nuclei. \citet{cui01} observed nuclei with absolute magnitude from $-17$ to $-21$ in $I$-band. The $I$-band luminosity of the identified putative nuclei is typically around a few percent of far-infrared luminosity of their host galaxy \citep{cui01}.

We estimated the absolute $I$-band magnitude of hypermassive star clusters formed in our simulations using {\tt P\'EGASE}. Some hypermassive star clusters have luminosity comparable to the observed putative nuclei. In the upper panels of Fig.~\ref{image}, there is one hypermassive star cluster with $M_{I}\sim -17.4$ in run $H_{\rm{PP}}$, and there are three hypermassive star clusters with $M_{I}\sim -17.7$, $-17.1$, and $-17.1$ in run $H_{\rm{TT}}$, respectively. These $I$-band absolute magnitudes are comparable to those of the galactic cores, which are $-18.5$ and $-17.5$ in run $H_{\rm{PP}}$ and $-17.7$ and $-15.7$ in run $H_{\rm{TT}}$. These results indicate  that the galactic nuclei and hypermassive star clusters are not distinguishable from their luminosities. The $I$-band (bolometric) luminosity of hypermassive star clusters is a few percent ($\gtrsim 10$\%) of the bolometric luminosity of the host galaxy with $\sim 10^{11}~L_{\odot}$. These characteristics are in good agreement with the observations.

We also compare the spatial distribution of the observed putative nuclei with
that of simulated hypermassive star clusters. If we select relatively
compact ULIRGs, the average separation between nuclei is about $1~\rm{kpc}$
\citep{cui01}.  In our simulations, the separation between hypermassive star
clusters and galactic cores depends on the evolution phase. However, the typical separation is a few $\rm{kpc}$ in the most luminous phase of the galaxies. Thus, the spatial distribution also agrees with the observations.

\subsection{Dust Extinction}

In this subsection, we investigate the effect of the dust extinction in $I$-band. Here, we use the high spatial resolution run, $H_{\rm{TT},5\rm{pc}}$, in which the softening length is $5~\rm{pc}$ for mock observatations, since the detailed structure of the ISM is important to estimate the extinction. In this run, similarly to run $H_{\rm{TT}}$, hypermassive star clusters with $\gtrsim 10^8~M_{\odot}$ form after the second encounter. The masses of the first, second, third most massive star clusters are $\sim 1.6 \times 10^8~M_{\odot}$,  $1.1 \times 10^8~M_{\odot}$, and $6.1 \times 10^7~M_{\odot}$, respectively.

The flux, taking dust extinction into account, from a star is estimated by $\displaystyle F = F_{0} \prod_{i} \exp(-\tau _i)$. Here, $F_{0}$ is the flux expected without dust extinction, and $\tau _i$ is optical depth for each SPH particle and $\tau_i = q_{\rm{d}}~ \sigma _{\rm{H},\rm{I}}~N_{\rm{H},i}~(Z_i/Z_{\odot})$ \citep{bek00,kob10}, where $q_{\rm{d}}$, $\sigma _{\rm{H},\rm{I}}$, $N_{\rm{H},i}$, and $Z_{i}$ represent the clumpiness parameter, the dust cross-section per H atom at $I$-band for the Milkyway dust model, column number density of atomic H of a SPH particle, and the metallicity of a SPH particle, respectively. The SPH particles between the star and the observer are used for calculating the optical depth, $\tau$. The clumpiness parameter is introduced in order to represent the clumpiness of the dust distribution in the ISM. It is assumed to be $0.15$, $0.5$, and $1.0$. The value of $q_{\rm{d}}=0.15$ is used in Ly$\alpha$ emitters \citep{kob10}, whereas that of $q_{\rm{d}}=1.0$ is the case without the dust clumpiness under the spatial resolution. Here, we do not estimate the ``clumpiness'' of SPH particles along the line of sight directly, since the ``real'' clumpiness comes from much smaller scale in the ISM. We adopt $\sigma _{\rm{H},\rm{I}}=4.0\times 10^{-22}~\rm{cm}^{2}$ \citep{dra03}. The column density of an SPH particle is given by $N_{\rm{H},i} = m_{i} W(r,h)$, where $m_{i}$ and $W(r,h)$ are the mass and 2-dimensional kernel function of the SPH particle, respectively, where $r$ is the 2-dimensional distance between the SPH and star particles in the projected plane, and $h$ is the kernel size of the SPH particle.

\begin{figure*}
\epsscale{1.0}
\plotone{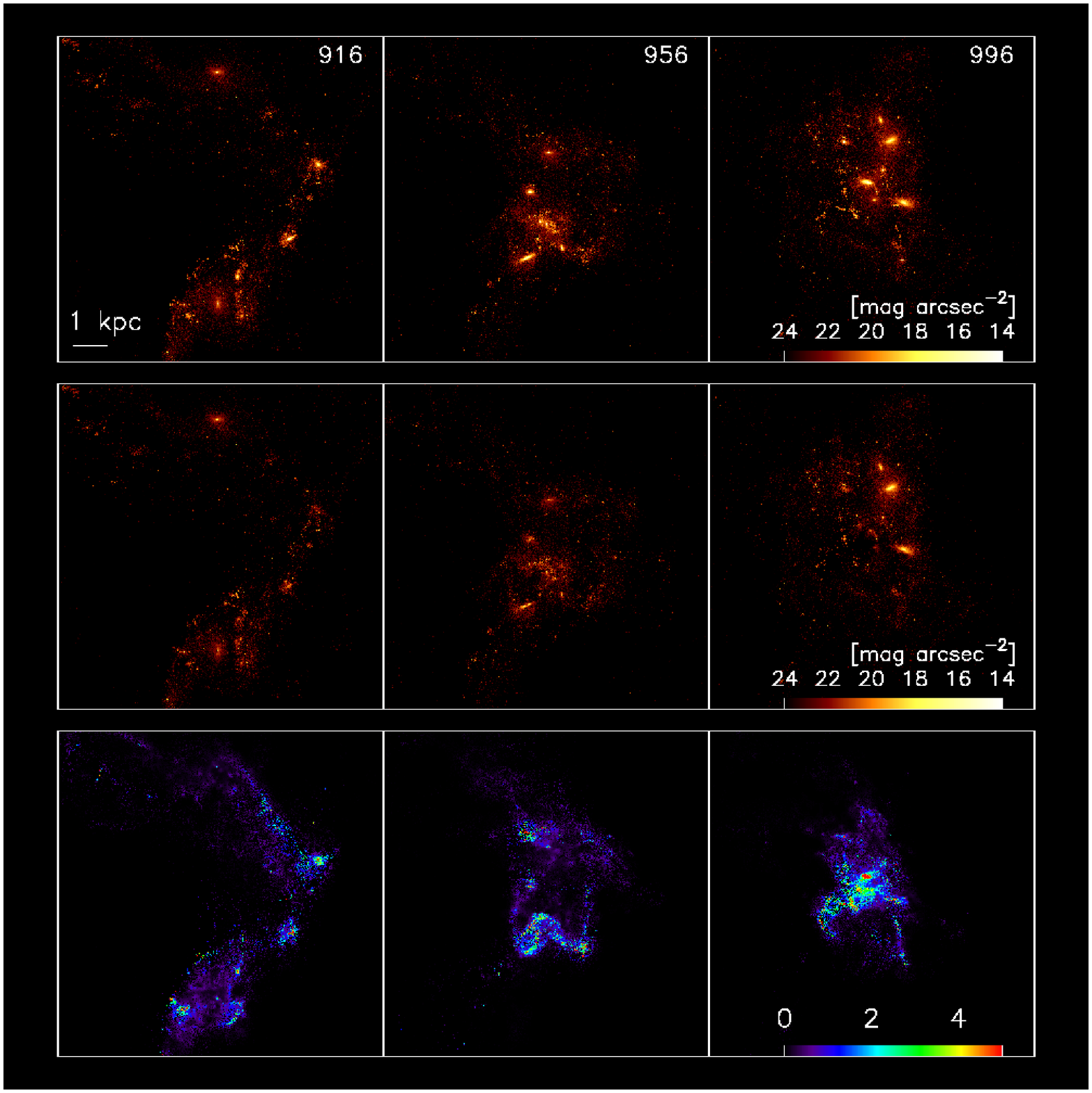}
\caption{
Time evolution of $I$-band map. The upper panels show the map without dust extinctions, and the middle panel shows the map taking dust extinction into account. Here, the clumpiness parameter is $q_d=0.5$. At the upper-right coner in all upper panels, simulation time ($\rm{Myr}$) is shown. The bottom panels show the amount of extinctions (the unit is magnitude).
\label{extinction_map}}
\end{figure*}

Figure~\ref{extinction_map} illustrates the effect of dust extinctions. The upper row shows the time evolution of the $I$-band map without extinction. The middle row represents the expected observation taking dust extinction with $q_d = 0.5$ into account for run $H_{\rm{TT},5\rm{pc}}$. Here, the map is projected to the orbital plane of the galaxies. After formation of hypermassive star clusters, strong dust extinctions take place in the clusters since there is a large amount of dusty gas. The dust extinction reduces the flux from the formed hypermassive star clusters. The extinction of the clusters in $I$-band is $A(I) \gtrsim 5$ magnitude in the $q_d=0.5$ case as shown in the bottom panels of Fig.~\ref{extinction_map}. In the models of $q_d=0.15$ and $q_d=1.0$, the extinctions are $A(I) \gtrsim 4$ and $A(I) \gtrsim 6$, respectively. The strong extinction continues till $970~\rm{Myr}$ (the left and the middle panels), which corresponds to the epoch about $50~\rm{Myr}$ after the formation of the clusters. In this dusty phase, multiple core structures are buried by dust. After $990~\rm{Myr}$ (the right panels), the clusters become optically thin because of consumption of dusty gas by star formation and escape of dusty gas from the cluster by SN feedback. As a result, the multiple cores appear. The extinction becomes $A_{I} \lesssim 1$ magnitude. On the other hand, the galactic central region is still buried by dust due to the continuous gas accretion.  In this phase, the extinction does not depend strongly on the clumpiness parameter since dusty gas is restricted to the central region of the cluster rather than distributed in the broad region of the cluster.

The timescale in which hypermassive star clusters become optically thin in $I$-band is less than the ULIRG lifetime ($\sim 200~\rm{Myr}$ as shown in Fig.~\ref{sfr}). Therefore, it is possible for the clusters to be observed as ULIRGs. Note, however, that the estimate of the dust extinction depends strongly on the distribution of the dust and that resolutions of current simulations are insufficient to resolve the ``real'' structure of the ISM. Futher high resolution simulations are necessary to involve the clumpiness of the dust distribution in the ISM.

\subsection{The Fraction of ULIRGs with Multiple Nuclei}

The fraction of ULIRGs with multiple nuclei strongly depends on the wavelength used for observation.  In the analysis of {\it HST} $I$-band data of $\sim 100$ ULIRGs samples by \cite{bor00} and \cite{cui01}, the fraction of the ULIRGs with multiple nuclei is $\sim 20\%$. In addition, \cite{cui01} have analyzed the nine samples of \cite{sur98} in $I$-band.  Two ULIRGs are found to have multiple nuclei, i.e., the fraction is consistent with the results of \cite{bor00} and \cite{cui01}.

On the other hand, \cite{bus02} and \cite{vei02} have claimed that the fraction of ULIRGs with multiple nuclei is less than $5\%$, based on their analysis of the data observed in near IR bands. Using the {\it HST} $H$-band data, \cite{bus02} reanalyzed 27 samples randomly selected from the 123 $I$-band samples, which include a part of Borne's and Cui's samples with multiple nuclei.  In their analysis, only one sample is classified as a ULIRG with multiple nuclei. The other sample includes some ULIRGs that have been already classified as ULIRGs with multiple nuclei by $I$-band image analysis \citep{bor00,cui01}. The reasons for excluding them from ULIRGs with multiple nuclei are as follows: (1) the nucleus seems to locate on the tidal arm and is thought to be the tidal arm or (2) multiple nuclei morphology do not appear in the $H$-band images.  Furthermore, \cite{vei02} analyzed the $K'$-band data of 118 ULIRGs taken by the Keck observatory. Their samples also include candidates of ULIRGs with multiple nuclei in HST $I$-band data. Some of them are excluded since multiple cores are not apparent in $K'$-band in spite of their appearance in $I$-band. As a result, only 5 ULIRGs ($\sim 4\%$) are classified as ULIRGs with multiple nuclei.

In order to understand the difference between the results obtained using images taken in different bands, we made $I$-, $H$-, and $K$-band images of our simulation data using {\tt P\'EGASE}. We assumed that galaxies were at $z \sim 0.1$ ($1'' \sim 2~{\rm kpc}$). The point spread function was assumed to be Gaussian, and its dispersion was calculated by $\sigma = \Delta x / \sqrt{2 \ln 2}$ , where $\Delta x$ is the spatial resolution. The spatial resolutions of $I$- and $H$-band images were given by the diffraction limit since we assumed the observations by {\it HST}. The spatial resolutions of $I$- and $H$-bands are $0''.1$ and $0''.2$, respectively. For $K$-band images, the seeing determines the spatial resolution since we assumed ground-based  observations by  the University of Hawaii $2.2~{\rm m}$ telescope or the Keck spectroscopy. The spatial resolution, therefore, is set to be $0''.5$ \citep{kim02}.  For comparison, we also made the emulated images with $K$-band by {\it James Web Space Telescope} ({\it JWST}) or ground-based $8$m telescopes with adaptive optics (AO).  We assume that the angular resolution of these future instruments is $\sim 0''.05$.

\begin{figure*}
\epsscale{1.0}
\plotone{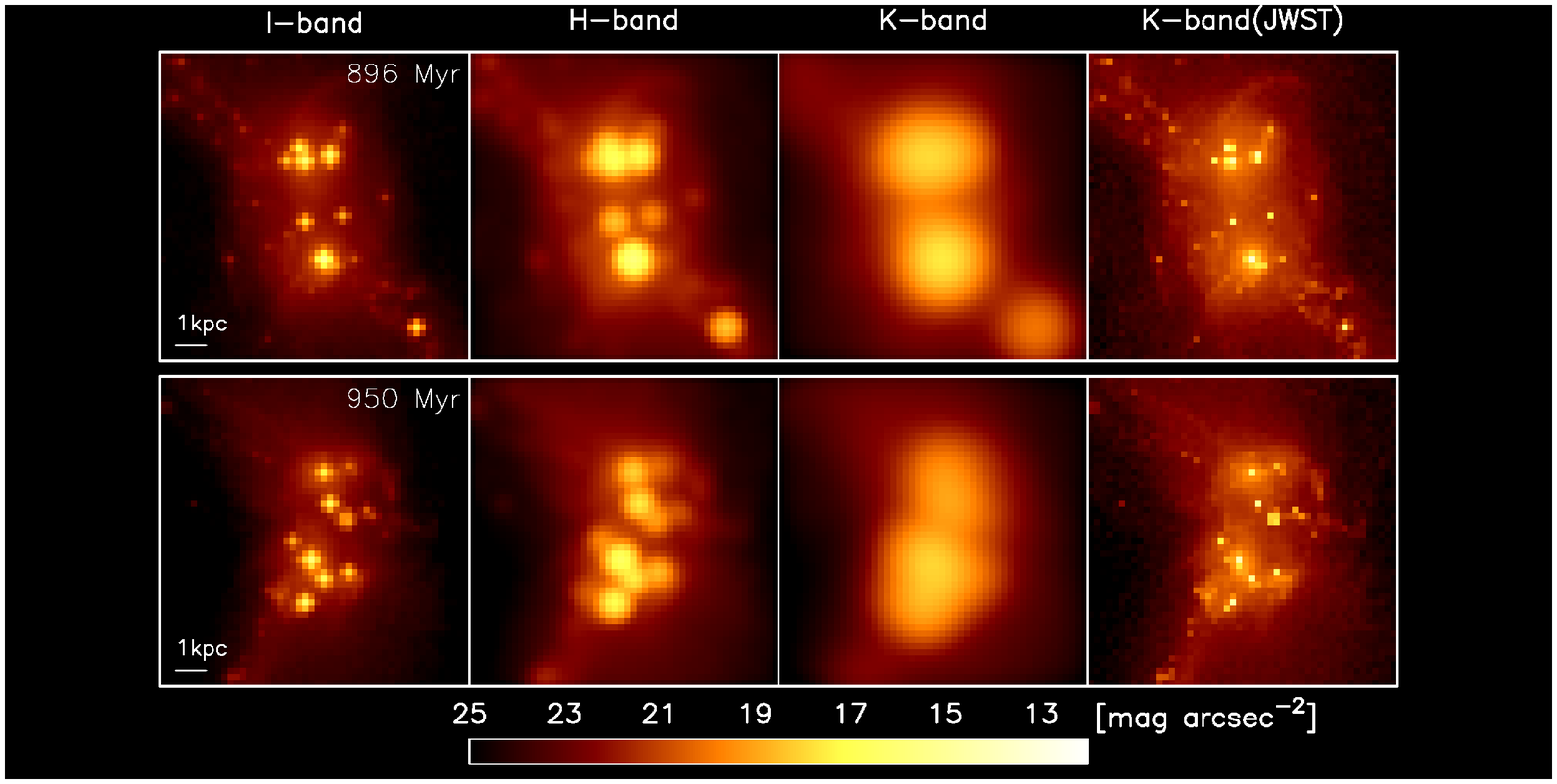}
\caption{
The observed images without dust extinction in our simulations. The upper panels show the images of run $H_{\rm{PP}}$ and bottom ones show those of run $H_{\rm{TT}}$. From left to right, we show images observed by $I$-, $H$-, and $K$-band with current instruments and $K$-bands with {\it JWST}.  The corresponding angular resolutions for these four bands are $0''.1$, $0''.2$, $0''.5$, and $0''.05$, respectively. See the text for details.
\label{multibands}}
\end{figure*}

Figure~\ref{multibands} shows $I$-, $H$-, and $K$-band images, in which dust extinction are not taken into account, of runs $H_{\rm{PP}}$ and $H_{\rm{TT}}$. In $I$-band images, multiple core-like structures are clearly visible, while such cores are blurred in other bands due to the limited resolution and the contribution of the luminosity of old stars.  In $H$-band images, cores are connected one another and form arm-like structures. As a result, the cores seem to be in arms. In $K$-band images, multiple core structures are not resolved at all.  As a result, ULIRGs seem to have only single or double cores. Thus, multiple cores were identified as a single or double cores in $H$- and $K'$-bands. Our analysis indicates that ULIRGs with multiple cores can be misclassified as ones with single or double cores. The apparent discrepancy in the observational fractions of ULIRGs with multiple nuclei might have been caused by this difference in the spatial resolution.

In the rightmost column in Fig.~\ref{multibands}, we show the emulated $K$-band images of our simulations with the angular resolution of $0''.05$. We can clearly see the multiple bright sources in these $K$-band images that were not visible in grand-base observations, since the resolution of the image is remarkably improved.  We expect that, in the near future, observations of (U)LIRGs by {\it JWST} or ground-based $8$m telescopes with AO will settle the problem of the band-to-band difference in the fraction of ULIRGs with multiple nuclei.

\begin{figure*}
\epsscale{1.0}
\plotone{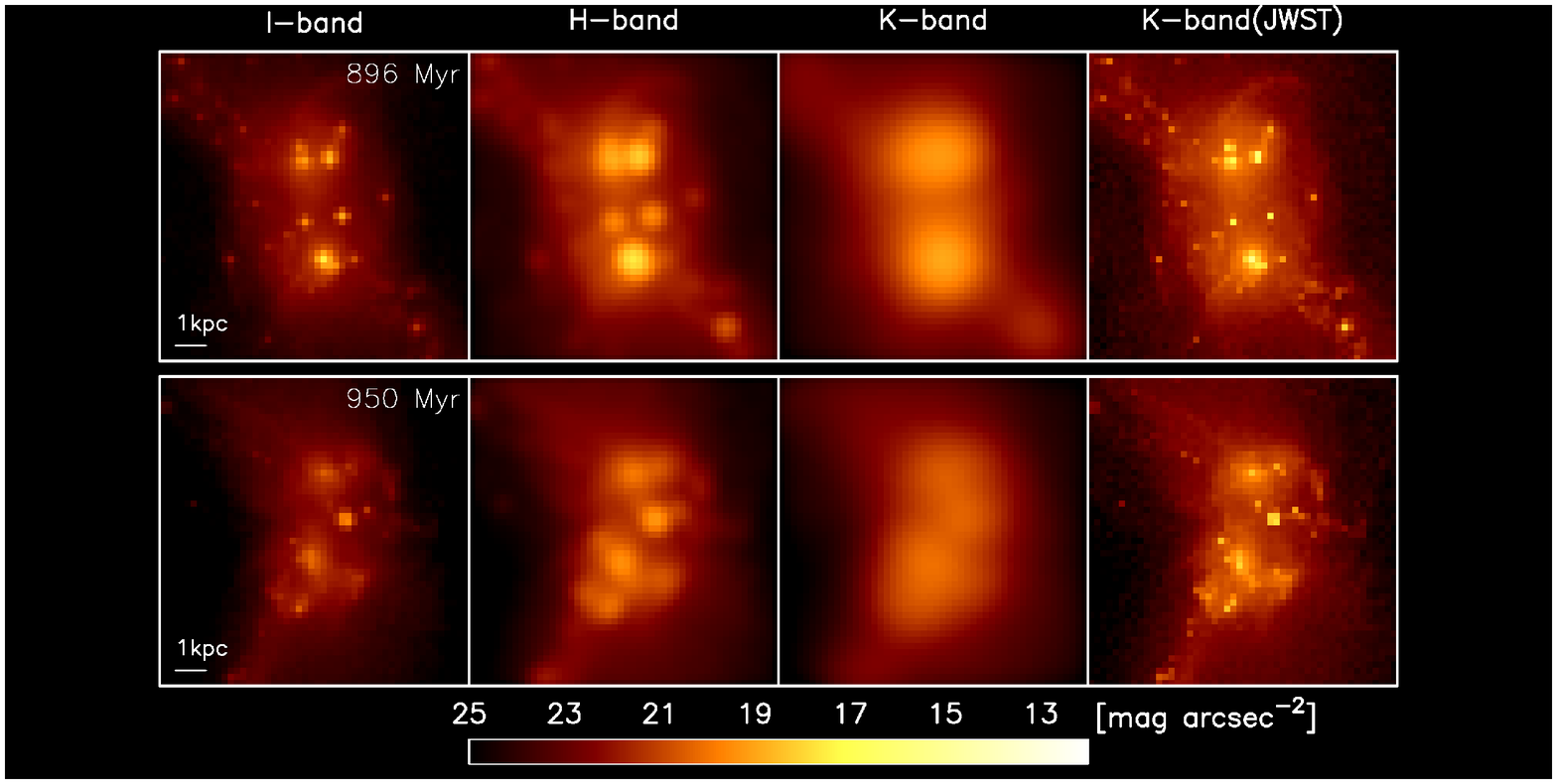}
\caption{
Same as Fig.~\ref{multibands} but taking dust extinction into account.
\label{multibands_dust}}
\end{figure*}

Figure \ref{multibands_dust} shows the observed $I$-, $H$-, and $K$-band images taking dust extinction into account. Here, the cross-sections per H atom at $H$- and $K$-bands are assumed to be $\sigma _{H,\rm{H}}=1.0\times 10^{-22}$ and $\sigma _{H,\rm{K}}=7.0\times 10^{-23}$, respectively \citep{dra03}. We adopt $q_d = 0.5$ as a fiducial value. In the $I$-band image, the hypermassive star clusters and the galactic cores are obscured because of dust extinction although the resolution is enough to observe them. On the other hand, in the $H$- and ground-based $K$-band images, dust extinction becomes weaker although the clusters and the galactic cores are not resolved. In the $K$-band image by {\it JWST} or ground-based $8$m telescopes with AO, the clusters and the galactic cores are clearly observable since not only dust extinction is small but also they are resolved. When dust extinction is significant in $I$-band, we need the observations by {\it JWST} or ground-based $8$m telescopes with AO.

\section{Conclusions and Discussion} \label{sec:Conclusions}

We have performed high resolution $N$-body/SPH simulations of merging galaxies. We found that hypermassive star clusters with $\sim 10^{8}~M_{\odot}$ form from disturbed gas disks in the central region ($\sim \rm{kpc}$). In these clusters, active star formations take place, so that some bright core structures appear in the merging galaxy. The features of formed hypermassive star clusters agree with the observations of ULIRGs with multiple nuclei. Although dust extinction may obscure the clusters after their formation, the clusters become optically thin within the ULIRG lifetime. These results indicate that one major merger of two spiral galaxies can explain complex structures of multiple nuclei in ULIRGs. Rare multiple major merger events on a short timescale are not necessary.

In the high-redshift universe, gas-rich major mergers must occur frequently. Our result suggests that these merging galaxies have multiple hypermassive star clusters that would be observed  as bright sources in either near or far infrared bands. Future observations using e.g., {\it JWST} or ground-based $8$m telescopes with AO are expected to find a number of such merging galaxies with multiple bright sources.

In our simulations, we assume that main feedback source is Type II SNe, and radiative feedback from newly formed stars is not taken into account. If the radiative feedback is effective, gas in hypermassive star clusters is heated and is ejected from the cluster \citep{kru10}. This effect may reduce the cluster mass. In order to study the effect, we need radiation-hydrodynamic simulations as performed by \citet{kru11}.

Since the hypermassive star clusters are compact ($\sim 10^{8} M_{\odot}$ within $20-30~\rm{pc}$), it is possible that star-star collisions and mergings occur in these clusters \citep{por99}. If the stellar mass loss in the main sequence phase is not very large, such merging of stars might result in the runaway growth of supermassive stars and the formation of intermediate-mass black holes (IMBHs) \citep{ebi01,por04}. These IMBHs are carried to the center of the galaxy together with the parent hypermassive star cluster by dynamical friction. If there is already a supermassive black hole at the center of the progenitor galaxies, the IMBH merges with the SMBH through dynamical friction and eccentricity evolution \citep{mat07}. If there was no supermassive black hole, multiple IMBHs conveyed to the center by the parent cluster finally merge each other to form the seed of a SMBH.

\acknowledgments

The authours thank the anonymous referee for many useful comments. The authors also thank Junichi Baba, Nozomu Kawakatu, Masakazu A. R. Kobayashi, and Tohru Nagao for helpful discussions. Numerical computations were carried out on Cray XT4 and GRAPE system at Center for Computational Astrophysics (CfCA) of National Astronomical Observatory of Japan. This project is supported by NEC Corporation, Molecular-Based New Computational Science Program of NINS, Grant-in-Aid for Scientific Research (17340059) of JSPS, and HPCI Strategic Program Field 5 ``The origin of matter and the universe''. T.R.S. is financially  supported by a Research Fellowship from the Japan Society for the Promotion of Science for Young Scientists. T.O. is financially supported by Grant-in-Aid for Scientific Research (S) by JSPS (20224002) and by Grant-in-Aid for Young Scientists (21840015).

\end{document}